\documentclass[aps,showpacs,twocolumn]{revtex4-1}
\usepackage{epsfig}
\usepackage{graphicx,color,dcolumn}
\usepackage{epstopdf}
\usepackage{amsmath,amssymb}
\usepackage{multirow}
\usepackage{diagbox}
\usepackage{changes}
\usepackage{booktabs}
\usepackage{threeparttable}
\usepackage{subfigure}
\usepackage{hyperref}

\newcommand{\ty}{Yue Tan}

\newcommand{\hhx}{Hongxia Huang}
\newcommand{\jlp}{Jialun Ping}

\newcommand{\nj}{Nanjing}

\begin{document}
\title{  Further study of $c\bar{c}c\bar{c}$ system within a chiral quark model }

\author{Yuheng Wu}
\email[E-mail: ]{191002007@njnu.edu.cn  }
\affiliation{Department of Physics, Yancheng Institute of Technology, Yancheng 224000, P. R. China}
\affiliation{Department of Physics, \nj~ Normal University, \nj~ 210023, P.R. China}

\author{Xuejie Liu}
\email[E-mail: ]{liuxuejie2023@htu.edu.cn  }
\affiliation{Department of Physics, Henan Normal University, Xinxiang 453007, P. R. China}

\author{\jlp}
\email[E-mail: ]{jlping@njnu.edu.cn}
\affiliation{Department of Physics, \nj~ Normal University, \nj~ 210023, P.R. China}
\date{\today}

\author{\ty}
\email[E-mail: ]{tanyue@ycit.edu.cn (Corresponding author)}
\affiliation{Department of Physics, Yancheng Institute of Technology, Yancheng 224000, P. R. China}

\author{\hhx}
\email[E-mail: ]{hxhuang@njnu.edu.cn (Corresponding author) }
\affiliation{Department of Physics, \nj~ Normal University, \nj~ 210023, P.R. China}

\begin{abstract}
Inspired by the recent ATLAS and CMS experiments on the invariant mass spectrum of $J/\psi J/\psi$, we systematically study the $c\bar{c}c\bar{c}$ system of $J^{P}=0^{+}$. In the framework of chiral quark model, we have carried out bound-state calculation and resonance-state calculation respectively by using Real-scaling method. The results of bound-state calculation show that there are no bound states in the $c\bar{c}c\bar{c}$ with $0^{+}$ system. The resonance-state calculation shows that there are four  possible stable resonances: $R(6920)$, $R(7000)$, $R(7080)$ and $R(7160)$. $R(6920)$ and $R(7160)$ are experimental candidates for $X(6900)$ and $X(7200)$, whose main decay channel is $J/\psi J/\psi$. It is important to note that the another major  decay channel of  $R(7160)$ is  $\chi_{c0} \chi_{c0} $, and  the  $\chi_{c0} \chi_{c0} $ is also the main decay channel of $R(7000)$, $R(7080)$. Therefore, we propose to search experimentally for these two predicted resonances in the $\chi_{c0} \chi_{c0}$ invariant mass spectrum.
\end{abstract}

\maketitle

\section{Introduction} \label{introduction}
In the conventional quark model~\cite{Gell-Mann:1964ewy}, hadrons are classified into two categories: meson ($q\bar{q}$) and baryon (qqq).
Moreover, it also allows for the existence of multiquark states, such as tetraquark ($q^{2}\bar{q}^{2}$), pentaquark ($q^{4}\bar{q}$), dibaryon ($q^{6}$), and so on, which are permitted by Quantum Chromodynamics(QCD). The  discovery of $X(3872)$~\cite{Belle:2003nnu}, in 2003, marked the beginning of a fascinating exploration into the realm of exotic hadrons. During the following two decades, numerous experimental reports revealed an abundant of charmed exotic states, which plays an important role in advancing our comprehension of the internal nature of multiquarks system.

In 2020, the LHCb Collaboration reported a new state $X(6900)$ with the significance greater than 5$\sigma$ in the di-$J/\psi$ mass distribution, using p-p collision data at center-of-mass energies of $\sqrt{s}=7,8$ and $13$ TeV~\cite{LHCb:2020bwg}.
Assuming no interference with the nonresonant single-parton scattering (NRSPS) continuum, the mass and natural width of the narrow $X(6900)$ are $m=6905\pm11\pm7~\mbox{MeV}$ and $\Gamma=80\pm19\pm33~\mbox{MeV}$, respectively. When assuming the NRSPS continuum, the mass and width become $m=6886\pm11\pm11~\mbox{MeV}$ and $\Gamma=168\pm33\pm69~\mbox{MeV}$, respectively. In addition, a broad structure next to the di-$J/\psi$ mass threshold and a vague structures around $7.2$ GeV are observed, but neither their mass nor width is given.

Subsequently, CMS~\cite{CMS:2023owd} and ATLAS~\cite{ATLAS:2023bft} Collaborations confirmed the existence of $X(6900)$. In addition, some new resonances (R) are observed by CMS and ATALS Collaborations. The mass and decay width of these resonance structures from CMS are: 
\begin{eqnarray}
R_{0}:m&=&6552\pm10\pm12 ~\mbox{MeV}, ~~~\nonumber\\
~~\Gamma&=&124_{-23}^{+32}\pm33~~\mbox{MeV}; ~~~ \nonumber\\
R_{1}:m&=&6927\pm9\pm4 ~\mbox{MeV}, ~~~\nonumber\\
\Gamma&=&124_{-21}^{+24}\pm18~~\mbox{MeV}; ~~~ \nonumber\\
R_{2}:m&=&7287_{-18}^{+20}\pm5 ~\mbox{MeV}, ~~~\nonumber\\
\Gamma&=&95_{-40}^{+59}\pm19~~\mbox{MeV}. ~~~
\end{eqnarray}
While for ATLAS, in the di-$J/\psi$ and $J/\psi$$\psi(2S)$ channels, several resonances are:
\begin{eqnarray}
R_{0}:m&=&6.41\pm0.08_{-0.03}^{+0.08} ~~\mbox{GeV}, ~~~\nonumber\\
~~\Gamma&=&0.59\pm0.35_{-0.20}^{+0.12}~~\mbox{GeV}; ~~~ \nonumber\\
R_{1}:m&=&6.63\pm0.05_{-0.01}^{+0.08} ~~\mbox{GeV}, ~~~\nonumber\\
\Gamma&=&0.35\pm0.11_{-0.04}^{+0.11}~~\mbox{GeV}; ~~~ \nonumber\\
R_{2}:m&=&6.86\pm0.03_{-0.02}^{+0.01} ~~\mbox{GeV}, ~~~\nonumber\\
\Gamma&=&0.11\pm0.05_{-0.01}^{+0.02}~~\mbox{GeV}; ~~~\nonumber\\
R_{3}:m&=&7.22\pm0.03_{-0.04}^{+0.01} ~~\mbox{GeV}, ~~~\nonumber\\
\Gamma&=&0.09\pm0.06_{-0.05}^{+0.06}~~\mbox{GeV}; ~~~
\end{eqnarray}

On the theoretical side, the $cc\bar{c}\bar{c}$ states have been studied by many theorists with various methods before the experiment, such as MIT bag model~\cite{Heller:1985cb}, nonrelativistic quark model~\cite{Debastiani:2017msn,Lundhammar:2020xvw,Wang:2019rdo}, constituent quark model~\cite{Barnea:2006sd,Richard:2017vry}, color-magnetic interaction model~\cite{Wu:2016vtq}, potential model~\cite{Liu:2019zuc}, quark-gluon model~\cite{Chao:1980dv}, QCD-sum rule~\cite{Wang:2017jtz,Wang:2018poa,Chen:2016jxd,Chen:2018cqz}, and so on~\cite{Berezhnoy:2011xn,Karliner:2016zzc,Lloyd:2003yc,Ader:1981db}.

After the experimental reports, the study of fully-charm tetraquark states arose again. In Ref.~\cite{Jin:2020jfc}, Jin et al. employed two models, quark delocalization color screening model and chiral quark model, to study the fully-charm tetraquark system, and the results are qualitative consistent in two quark models. The $X(6900)$ state was explained as a $S$-wave compact resonance state with $J^{P}=0^{+}$. Besides, the authors also found resonance structures around the energies $6270$ MeV with $J^{P}=0^{+}$ and $7260$ MeV with $J^{P}=0^{+}$ and $1^{+}$, respectively. However, no signal was found for $X(6600)$ state. Li et al.~\cite{Li:2021ygk} applied the Bethe-Salpeter equation to investigate the states of fully charm tetraquark, and they obtained three ground states of $cc\bar{c}\bar{c}$ in the mass range from $6.4$ GeV to $6.5$ GeV. For $X(6900)$, they suggested that it is less likely to be the ground state of a compact $cc\bar{c}\bar{c}$ tetraquark, but may be the first or second excited states. In Ref.~\cite{Wang:2022yes}, Wang et al. calculated the $S$-wave $cc\bar{c}\bar{c}$ tetraquark states in the framework of nonrelativistic quark model. Their results show that in the $S$-wave sector, no resonance is found at the energy region of the $X(6200)$ and $X(6600)$ states. Although for $X(6900)$ they obtained masses with $J^{PC}=0^{++}$ and $2^{++}$ obove the experimental $100$ MeV, the decay widths are in agreement with the experiment. Moreover, the higher $0^{++}$ and $2^{++}$ resonances are found at about $7.2$ GeV with the widths of $60.6$ MeV and $91.2$ MeV.  
In Ref.~\cite{Wu:2024euj}, Wu et al. investigated the $S$-wave fully heavy tetraquark systems by utilizing the Gaussian expansion method. They found $X(6900)$ and $X(7200)$ can be explained as $0^{++}$ and $2^{++}$ systems. However, $X(6400)$ and $X(6600)$ were not seen in their calculation.  

For the fully-heavy tetraquark states, the theoretical explanations are far from convergent. By using QCD sum rule, Ref.~\cite{Chen:2020xwe} argued that the $X(6900)$ can be considered as a $P$-wave $cc\bar{c}\bar{c}$ tetraquark with $J^{PC}=0^{-+}$ or $1^{-+}$. In Ref.~\cite{Zhu:2020xni}, in the framework of Bethe-Salpeter and Regge trajectory relation, the author pointed out that the $X(6900)$ can be explained as a radially excited state with $J^{PC}=0^{++}(3S)$ or $J^{PC}=2^{++}(3S)$ or an orbitally excited $2P$ state. Deng et al. \cite{Deng:2020iqw} pointed out that the structure $X(6900)$ can be identified as the excited $[cc][\bar{c}\bar{c}]$ with $L=1~(L=2)$ in the color flux-tube model. In Ref.~\cite{Liu:2021rtn,liu:2020eha}, the authors suggested that $X(6900)$ may be caused by $1P$-, or $2S$-, or $1D$-wave states while the $X(7200)$ is the high-lying $1D$-wave states with $J^{PC}=0^{++},1^{++},2^{++},3^{++}$, or $4^{++}$. Dong et al. \cite{Dong:2022sef} interpret the $X(6200)$, $X(6600)$, $X(6900)$ and $X(7200)$ structures as $1S$-wave, $1P/2S$-wave, $1D/2P$-wave and $2D/3P/4S$-wave fully charmed tetraquark state, respectively.
In Ref.~\cite{Yu:2022lak}, the authors argue that $X(6600)$ can be explained as $1P$ states with $IJ^{P}=0^{-+}$, $1^{-+}$ and $2^{-+}$. The possible assignments for $X(6900)$ can be $0^{++}(2S)$, $2^{++}(2S)$ or $0^{-+}(1P)$ states. $X(7200)$ can be interpreted as $2P$ states with $IJ^{P}=0^{-+}$, $1^{-+}$ and $2^{-+}$. In Ref.~\cite{Ortega:2023pmr}, P. G. Ortega et al. find two candidates for $X(6200)$, one for $X(6600)$, two for $X(6700)$ and four for $X(7200)$ tetraquarks.
In Refs.~\cite{Agaev:2023gaq,Agaev:2023wua,Agaev:2023ruu,Agaev:2023rpj}, S. S. Agaev et al. employed QCD sum rule calculated fully heavy tetraquark systems. From these works, we can see that the states composed of $\chi_{c0}$ and $\chi_{c1}$ play an important role in explaining the experimental discovery of $X(6600)$, $X(6900)$ and $X(7200)$. 
In Ref.\cite{Mutuk:2021hmi}, 
by employing a nonrelativistic model, H. Mutuk investigated fully heavy tetraquark systems with diquark-antidiquark structures. The results show that first radially excited states are considerably lower than their corresponding $(2S$-$2S)$ two-meson threshold.
More results and discussions for the $cc\bar{c}\bar{c}$ tetraquark system are given in Refs.~\cite{Yang:2021hrb,Lu:2020cns,Yang:2020wkh,Liang:2021fzr,Dong:2020nwy,Wan:2020fsk,Feng:2020riv,Maciula:2020wri,Feng:2020qee,
Goncalves:2021ytq,Huang:2021vtb,Becchi:2020uvq,Bedolla:2019zwg,Faustov:2020qfm,Gong:2020bmg,Gordillo:2020sgc,Karliner:2020dta,Ke:2021iyh,Weng:2020jao,
Zhang:2020xtb,Zhao:2020jvl,Zhao:2020nwy,Wu:2022qwd,Zhuang:2021pci,Wang:2022xja,Mutuk:2022nkw}.

In our previous work~\cite{Chen:2021crg}, in the framework of chiral quark model (ChQM), we systematically investigate the fully-charm tetraquark states. The narrow structure $X(6900)$ was explained as $S$-wave resonance states with quantum numbers $J^{PC}=0^{++},~2^{++}$ and many resonance states above $7.0$ GeV. Some of them can be explained as $X(7200)$. However, too many resonance states are obtained in the previous work, since only the central part of the quark-(anti)quark interactions is take into account and some resonance states that can decay into tetraquark channels composed of $\chi_{c0}$, $\chi_{c1}$ and $\chi_{c2}$ excited mesons are not considered.
In the present work, we still employ the ChQM to study the fully-charm tetraquark states. Unlike the previous work, here we consider not only the central interactions, but also add spin-orbit and tensor coupling. Moreover, tetraquark channels composed of $\chi_{c0}$, $\chi_{c1}$ and $\chi_{c2}$ excited mesons are added to search for the possible decay channel.

The structure of this paper is as follows. Section II gives a brief description of the quark model and wave functions. Section III is devoted to the numerical results and discussions. The summary is shown in the last section.

\section{Chiral quark model, wave function of $c\bar{c}c\bar{c}$ system} \label{wavefunction and chiral quark model}
\subsection{Chiral quark model}
The chiral quark model has been applied successfully in describing the hadron spectra and hadron-hadron interactions.
The details of the model can be found in Refs. \cite{Vijande:2004he,Hu:2021nvs,Tan:2020ldi}. In this writing, we introduce a chiral quark model with scalar nonet exchange.
The Hamiltonian of the chiral quark model is given as follows,
\begin{eqnarray}
\label{H}\nonumber
H &=&\sum_{i=1}^4 m_i +\frac{\vec{p}_{12}^2}{2\mu_{12}} +\frac{\vec{p}_{34}^2}{2\mu_{34}}+\frac{\vec{p}_{1234}^2}{2\mu_{1234}} \\
&+&  \sum_{i<j=1}^n [ V_{con}(r_{ij}) +V_{oge}(r_{ij}) ],
\end{eqnarray}
where $m_i$ is the constituent mass of $i$-th quark (antiquark), and $\mu$ is the reduced masse of two interacting quarks or quark-clusters. In Eq. \ref{H}., the specific forms of momentum and reduced mass can be written as follows.
\begin{subequations}
\label{moment}
\begin{align}
\mu_{12}&=\frac{m_{{1}}  m_{{2}}}{m_{{1}} + m_{{2}}}, \mu_{34}=\frac{m_{{3}}  m_{{4}}}{m_{{3}} + m_{{4}}}   \\
\mu_{1234}&=\frac{(m_1+m_2)(m_3+m_4)}{m_1+m_2+m_3+m_4},\\
p_{12}&=\frac{m_2p_1-m_1p_2}{m_1+m_2},p_{34} = \frac{m_4p_3-m_3p_4}{m_3+m_4}  \\
p_{1234}&=\frac{(m_3+m_4)p_{12}-(m_1+m_2)p_{34}}{m_1+m_2+m_3+m_4}.
\end{align}
\end{subequations}

The different terms of the potential contain central, spin-orbit contributions and tensor force. However, it's difficult for us to deal with spin-orbit and tensor force contributions when  it comes to the computation  of four quarks. Thus, spin-orbit coupling, tensor force and central force would contribute to our two-quark calculation while only central force would contribute to our four quark calculation.

 $V_{con}(r_{ij})$ is the confining potential, mimics the ``confinement" property of QCD. The $V_{con}(r_{ij})$ term includes central force $V_{con}^{C}(r_{ij})$ and spin-orbit force $V_{con}^{SO}(r_{ij})$.

\begin{subequations}
\label{confinement}
\begin{align}
    V_{con}^C(r_{ij}) &= ( -a_{c} r_{ij}^{2}-\Delta) \boldsymbol{\lambda}_i^c \cdot \boldsymbol{\lambda}_j^c\\ \nonumber
    V_{con}^{SO}(r_{ij}) &= -\boldsymbol{\lambda}_i^c \cdot \boldsymbol{\lambda}_j^c \frac{a_c}{4m_i^2m_j^2}[ ((m_i^2+m_j^2)(1-2a_s) \\ \nonumber
&+ 4m_im_j(1-a_s))(\vec{S}_{+}\cdot \vec{L}) \\
&+((m_j^2-m_i^2)(1-2a_s))(\vec{S}_{-}\cdot \vec{L}) ]\\ \nonumber
\end{align}
\end{subequations}

The second potential $V_{oge}(r_{ij})$ is one-gluon exchange interaction reflecting the ``asymptotic freedom" property of QCD. The $V_{oge}(r_{ij})$ term contains central force $V_{oge}^{C}(r_{ij})$, spin-orbit force $V_{oge}^{SO}(r_{ij})$ and tensor force $V_{oge}^{T}(r_{ij})$.

\begin{subequations}
\label{oge}
\begin{align}
    V_{oge}^C(r_{ij}) &=\frac{\alpha_{cc}}{4} \boldsymbol{\lambda}_i^c \cdot \boldsymbol{\lambda}_{j}^c
\left[\frac{1}{r_{ij}}-\frac{1}{6m_im_jr_0^2}\boldsymbol{\sigma}_i\cdot
\boldsymbol{\sigma}_j \frac{e^{-\frac{r_{ij}}{r_{0}} }}{r_{ij}}\right]   \\ \nonumber
    V_{oge}^{SO}(r_{ij}) &= -\frac{1}{16} \frac{\alpha_{cc}\boldsymbol{\lambda}_i^c \cdot \boldsymbol{\lambda}_j^c}{4m_i^2m_j^2}[\frac{1}{r_{ij}^3}-\frac{e^{-r_{ij}/r_g(\mu)}}{r_{ij}^3}(1+\frac{r_{ij}}{r_g(\mu)})] \\ \nonumber
    &[ (m_i^2+m_j^2+4m_im_j)(\vec{S}_{+}\cdot \vec{L})\\
    &+(m_j^2-m_i^2)(\vec{S}_{-}\cdot \vec{L}) ]\\ \nonumber
V_{oge}^{T}(r_{ij}) &= -\frac{1}{16} \frac{\alpha_{cc}\boldsymbol{\lambda}_i^c \cdot \boldsymbol{\lambda}_j^c}{4m_i^2m_j^2}[\frac{1}{r_{ij}^3}-\frac{e^{-r_{ij}/r_g(\mu)}}{r_{ij}}(\frac{1}{r_{ij}^2}\\
&+\frac{1}{3r^2_g(\mu)} +\frac{1}{r_{ij}r_g(\mu)})]S_{ij} \\ \nonumber
\end{align}
\end{subequations}

$\boldsymbol{\sigma}$ are the $SU(2)$ Pauli matrices; $\boldsymbol{\lambda}_{c}$ are $SU(3)$ color Gell-Mann matrices,
$r_{0}(\mu_{ij})=\frac{r_0}{\mu_{ij}}$ and $\alpha_{cc}$ is a coupling constant between $c(\bar{c})$ quarks.

All the parameters are determined by fitting the meson spectrum, taking into account only a quark-antiquark component. They are shown in Table~\ref{modelparameters}. The calculated masses of the mesons involved in the present work are shown in Table~\ref{mesonmass}.

\begin{table}[t]
\begin{center}
\caption{The quark model parameters.\label{modelparameters}}
\begin{tabular}{ccccccccccc}
\hline\hline\noalign{\smallskip}
$m_{c}$(MeV)&  $a_{c}$(MeV) &$\Delta$(MeV)   &$\hat{r}_0$(MeV) &$\hat{r}_g$(MeV)   \\
        4978&            98 &         -18.1           &81.0             &100.6    \\
$\alpha_{cc}$ & $a_{s}$ \\
0.56          & 0.77 \\
\hline\hline
\end{tabular}
\end{center}
\end{table}

\begin{table}[]
\caption{ \label{mesonmass}  Numerical results for the this work ( with a  harmonic form confinement ), the ChQM2( with a  color screening form confinement ) and LP model. (unit: MeV).\label{mesons}}
\begin{tabular}{ccccccc}
\hline\hline\noalign{\smallskip}
    Meson         &    This work    & ChQM2\cite{Vijande:2004he} &    LP\cite{Deng:2016stx} & EXP.(PDG) \\ \hline
    $\eta_{c}$    &    2980         &  2990 &   2983 & 2983.9$\pm$0.4\\
    $\eta_{c}(2S)$&    3637         &  3627 &   3635 & 3637.7$\pm$1.1\\
    $\eta_{c}(3S)$&    4132         &   -   &   4048 &               -\\
    $J/\psi$      &    3100         &  3097 &   3097 & 3096.9$\pm$0.006\\
    $\psi(2S)$    &    3713         &  3685 &   3679 & 3686.1$\pm$0.06\\
    $h_c$         &    3508         &  3507 &   3522 & 3525.37$\pm$0.14\\
    $\chi_{c0}$   &    3428         &  3436 &   3415 & 3414.71$\pm$0.30\\
    $\chi_{c1}$   &    3493         &  3494 &   3516 & 3510.67$\pm$0.05\\
    $\chi_{c2}$   &    3543         &  3526 &   3552 & 3556.17$\pm$0.07\\
\hline\hline
\end{tabular}
\end{table}

\subsection{The wave function of $c\bar{c}c\bar{c}$ system}
There are two physically important structures, dimeson ($c\bar{c}$-$c\bar{c}$) and diquark ($cc$-$\bar{c}\bar{c}$)  in the fully-charm system, which are considered in the present calculation. The wave functions of every structure all consists of four parts: orbital, spin, flavor and color. The flavor and color wave functions are independent, while the orbital and spin wave functions need to be coupled together first.

In the first step, we couple the orbital wave function $|\psi_{L_1,mL_1} \rangle$ and spin wave function $| \chi_{S_1,mS_1}^{\sigma} \rangle$ of the first cluster to obtain $|\psi_{J_1,mJ_1} \rangle$.
\begin{align}
 |\psi_{J_1,mJ_1}\rangle =  |\psi_{L_1,mL_1} \rangle \otimes  |\chi_{S_1,mS_1}^{{\sigma}} \rangle.
\end{align}
Similarly, the orbital wave function $|\psi_{L_2,mL_2} \rangle$ and spin wave function $| \chi_{S_2,mS_2}^{\sigma} \rangle$ of the second cluster are coupled to yield $|\psi_{J_2,mJ_2} \rangle$.
\begin{align}
 |\psi_{J_2,mJ_2}\rangle =  |\psi_{L_2,mL_2} \rangle \otimes  |\chi_{S_2,mS_2}^{{\sigma}} \rangle.
\end{align}
 The wave functions of the two clusters are then coupled together to form the wave function $|\psi_{J_{12},mJ_{12}} \rangle$.
 \begin{align}
 |\psi_{J_{12},mJ_{12}}\rangle =  |\psi_{J_1,mJ_1}\rangle \otimes  |\psi_{J_2,mJ_2}\rangle.
\end{align}
 Finally, this wave function $|\psi_{J_{12}} \rangle$ is coupled to the orbital wave function $|\psi_{L_3,mL_3} \rangle$ between the two clusters to form the total orbit-spin wave function $|\psi_{i}^{LS}\rangle$.
 \begin{align}
 &|\psi_{i}^{LS}\rangle =|\psi_{J_{12},mJ_{12}}\rangle  \otimes  |\psi_{L_3,mL_3} \rangle.  \\
 &i \equiv [J_1,J_2,J_{12},L_3].
 \end{align}
 In this paper, we focus on the $0^{+}$ system, considering a total of 7 combinations for $[J_1, J_2, J_{12}, L_3]$ under the fixed inter-cluster relative motion in the $S$-wave ($L_3=0$), as illustrated in TABLE \ref{measons}.

 Considering that the difference of each structure is particle order, we only present the spatial wave-function, spin wave-function and color wave-function of the dimeson structure and the diquark structure in the next subsection for simplicity.

\begin{table}[]
\caption{ \label{LScouple}  Different combinations of J-J coupling.\label{measons}}
\begin{tabular}{cccccccc}
\hline\hline\noalign{\smallskip}
\multicolumn{2}{l}{$L_3=0$} & \multicolumn{2}{c}{$J_{1}=0$} &~~~~&\multicolumn{2}{c}{$J_{2}=0$}& index(i)   \\
                                     && $L_1=0$ & $S_1=0$            & & $L_2=0$ & $S_2=0$           & 1       \\
                                     && $L_1=1$ & $S_1=1$            & & $L_2=1$ & $S_2=1$           & 2       \\
                                     \cline{3-7}
                                     && \multicolumn{2}{c}{$J_{1}=1$}& &\multicolumn{2}{c}{$J_{2}=1$}& index(i)   \\
                                     && $L_1=0$ & $S_1=1$             && $L_2=0$ & $S_2=1$           & 3       \\
                                     && $L_1=1$ & $S_1=0$             && $L_2=1$ & $S_2=0$           & 4       \\
                                     && $L_1=1$ & $S_1=1$             && $L_2=1$ & $S_2=1$           & 5       \\
                                     && $L_1=1$ & $S_1=1$             && $L_2=1$ & $S_2=0$           & 6       \\ \hline
                                     && \multicolumn{2}{c}{$J_{1}=2$}& &\multicolumn{2}{c}{$J_{2}=2$}& index(i)   \\
                                     && $L_1=1$ & $S_1=1$             && $L_2=1$ & $S_2=1$           & 7       \\
\hline
\end{tabular}
\end{table}

\subsubsection{spatial wave function}

In GEM \cite{Hiyama:2003cu}, the radial part of the orbital wave function is expanded by a set of Gaussians:
\begin{subequations}
\label{radialpart}
\begin{align}
\psi_L(\mathbf{r}) & = \sum_{n=1}^{n_{\rm max}} c_{n}\psi^G_{nlm}(\mathbf{r}),\\
\psi^G_{nlm}(\mathbf{r}) & = N_{nl}r^{l}
e^{-\nu_{n}r^2}Y_{lm}(\hat{\mathbf{r}}),
\end{align}
\end{subequations}
where $N_{nl}$ are normalization constants,
\begin{align}
N_{nl}=\left[\frac{2^{l+2}(2\nu_{n})^{l+\frac{3}{2}}}{\sqrt{\pi}(2l+1)}
\right]^\frac{1}{2}.
\end{align}
$c_n$ are the variational parameters, which are determined dynamically. The Gaussian size parameters are chosen according
to the following geometric progression
\begin{equation}\label{gaussiansize}
\nu_{n}=\frac{1}{r^2_n}, \quad r_n=r_1a^{n-1}, \quad
a=\left(\frac{r_{n_{\rm max}}}{r_1}\right)^{\frac{1}{n_{\rm max}-1}}.
\end{equation}
This procedure enables optimization of the using of Gaussians, as small as possible Gaussians are used.

\subsubsection{spin wave function}
Because of no difference between spin of quark and antiquark, the meson-meson structure has the same spin wave function as
the diquark-antidiquark structure. The spin wave functions of the sub-cluster are shown below.
\begin{align}
&\chi_{11}^{\sigma}=\alpha\alpha,~~
\chi_{10}^{\sigma}=\frac{1}{\sqrt{2}}(\alpha\beta+\beta\alpha),~~
\chi_{1-1}^{\sigma}=\beta\beta,\nonumber \\
&\chi_{00}^{\sigma}=\frac{1}{\sqrt{2}}(\alpha\beta-\beta\alpha),
\end{align}

\subsubsection{flavor wave function\label{sec_flavor}}

We have two flavor wave functions of the system,
\begin{align}\label{sec}
|F_{1}\rangle&= c_1\bar{c}_2c_3\bar{c}_4 \nonumber \\
|F_{2}\rangle&= c_1c_3\bar{c}_2\bar{c}_4
\end{align}
$|F_{1}\rangle$ is for meson-meson structure, and $|F_{2}\rangle$ is for diquark-antidiquark structure.

\subsubsection{color wave function}
The colorless tetraquark system has four color wave functions, two for meson-meson structure, $1\otimes1$ ($C_1$), $8\otimes8$ ($C_2$),
and two for diquark-antidiquark structure, $\bar{3}\otimes 3$ ($C_3$) and $6\otimes \bar{6}$ ($C_4$).

\begin{align}
&|C_{1}\rangle =  \sqrt{\frac{1}{9}} ( r_1\bar{r}_2r_3\bar{r}_4+r_1\bar{r}_2g_3\bar{g}_4+r_1\bar{r}_2b_3\bar{b}_4+g_1\bar{g}_2r_3\bar{r}_4 \nonumber\\
& +g_1\bar{g}_2g_3\bar{g}_4 +g_1\bar{g}_2b_3\bar{b}_4+b_1\bar{b}_2r_3\bar{r}_4  +b_1\bar{b}_2g_3\bar{g}_4+b_1\bar{b}_2b_3\bar{b}_4)  \nonumber\\
&|C_{2}\rangle = \sqrt{\frac{1}{72}}(3r_1\bar{b}_2b_3\bar{r}_4+3r_1\bar{g}_2g_3\bar{r}_4+3g_1\bar{b}_2b_3\bar{g}_4                   \nonumber\\
&+3b_1\bar{g}_2g_3\bar{b}_4+3g_1\bar{r}_2r_3\bar{g}_4+3b_1\bar{r}_2r_3\bar{b}_4+2r_1\bar{r}_2r_3\bar{r}_4 \nonumber\\
&+2g_1\bar{g}_2g_3\bar{g}_4+2b_1\bar{b}_2b_3\bar{b}_4 -r_1\bar{r}_2g_3\bar{g}_4-g_1\bar{g}_2r_3\bar{r}_4\nonumber\\
&-b_1\bar{b}_2g_3\bar{g}_4-b_1\bar{b}_2r_3\bar{r}_4-g_1\bar{g}_2b_3\bar{b}_4-r_1\bar{r}_2b_3\bar{b}_4). \nonumber\\
&|C_{3}\rangle = \sqrt{\frac{1}{12}}(r_1g_3\bar{r}_2\bar{g}_4-r_1g_3\bar{g}_2\bar{r}_4+g_1r_3\bar{g}_2\bar{r}_4-g_1r_3\bar{r}_2\bar{g}_4 \nonumber \\
&+r_1b_3\bar{r}_2\bar{b}_4-r_1b_3\bar{b}_2\bar{r}_4+b_1r_3\bar{b}_2\bar{r}_4-b_1r_3\bar{r}_2\bar{b}_4+g_1b_3\bar{g}_2\bar{b}_4 \nonumber \\
&-g_1b_3\bar{b}_2\bar{g}_4+b_1g_3\bar{b}_2\bar{g}_4-b_1g_3\bar{g}_2\bar{b}_4). \nonumber \\
&|C_{4}\rangle = \sqrt{\frac{1}{24}}(2r_1r_3\bar{r}_2\bar{r}_4+2g_1g_3\bar{g}_2\bar{g}_4+2b_1b_3\bar{b}_2\bar{b}_4+r_1g_3\bar{r}_2\bar{g}_4 \nonumber \\
&    +r_1g_3\bar{g}_2\bar{r}_4 +g_1r_3\bar{g}_2\bar{r}_4+g_1r_3\bar{r}_2\bar{g}_4+r_1b_3\bar{r}_2\bar{b}_4 +r_1b_3\bar{b}_2\bar{r}_4\nonumber \\
&+b_1r_3\bar{b}_2\bar{r}_4 +b_1r_3\bar{r}_2\bar{b}_4+g_1b_3\bar{g}_2\bar{b}_4+g_1b_3\bar{b}_2\bar{g}_4+b_1g_3\bar{b}_2\bar{g}_4\nonumber \\
&+b_1g_3\bar{g}_2\bar{b}_4).\nonumber \\
\end{align}

\subsubsection{total wave function}

Finally, the total wave function of the tetraquark  system is written as:
\begin{equation}
\Psi_{JM_J}^{i,j,k}={\cal A} \psi_{i}^{LS}  F_j C_k,
\end{equation}
where the $\cal{A}$ is the antisymmetry operator of the system which guarantees the antisymmetry of the total wave functions when identical particles exchange. Since the four particles in $c\bar{c}c\bar{c}$ system are all identical, the antisymmetry operator can be written as:
\begin{align}
\mathcal{A} = 1 - P(13)-P(24)+P(1324).
\end{align}
At last, we solve the following Schr\"{o}dinger equation to obtain eigen-energies of the system, with the help of the Rayleigh-Ritz variational
principle.
\begin{equation}
H\Psi_{JM_J}^{i,j,k}=E\Psi_{JM_J}^{i,j,k},
\end{equation}
where $\Psi_{JM_J}^{i,j,k}$ is the wave function of the four-quark states, which is the linear combinations of the above channel wave functions.

\section{Real-scaling method}

In the quark model, each colorful configuration, including  both the color octet and the diquark structure, manifests as a resonant state due to color attraction. Consequently, the quark model predicts an abundance of resonant states, including false resonant states, surpassing the number experimentally observed. Actually, the excess of resonant states arises due to calculation dimension limitations and leads to the inclusion of false resonances. To address this challenge, we have adopted the Real-Scaling method. In this approach, the expansion width $R$ of the orbital wave function between two subgroups is multiplied by a scaling factor $\alpha$, denoted as $R \rightarrow \alpha R$. As $\alpha$ increases, the calculating space expands, causing the energy of false resonance states to progressively decrease until they decay to the corresponding threshold channel. Conversely, genuine resonance states remain stable throughout this process. The stability of genuine resonances is characterized by two scenarios:

(1) Weak Coupling: When the energy of a scattering state significantly differs from that of the resonance, signifying weak or no coupling between the resonances and scattering states, the resonance manifests as a stable straight line, as depicted in Fig. 1(a).

(2) Strong Coupling: When the energy of a scattering state converges with that of the resonance, indicating strong coupling, an avoid crossing structure emerges between two declining lines, as illustrated in Fig. 1(b).

The decay width can be estimated from the slopes of the resonance and scattering states using Eq.~(\ref{formula_RSM}), where $S_r$ denotes the slope of the resonance, $S_s$ denotes the slope of the scattering state, and $\alpha_c$ represents the energy level difference between the resonance and the scattering state. Furthermore, as $\alpha$ increases continually, the avoid crossing structure repeats, providing valuable insights into the resonance behavior.

\begin{figure}[htp]
  \setlength {\abovecaptionskip} {-0.1cm}
  \centering
  \resizebox{0.50\textwidth}{!}{\includegraphics[width=3cm,height=2.2cm]{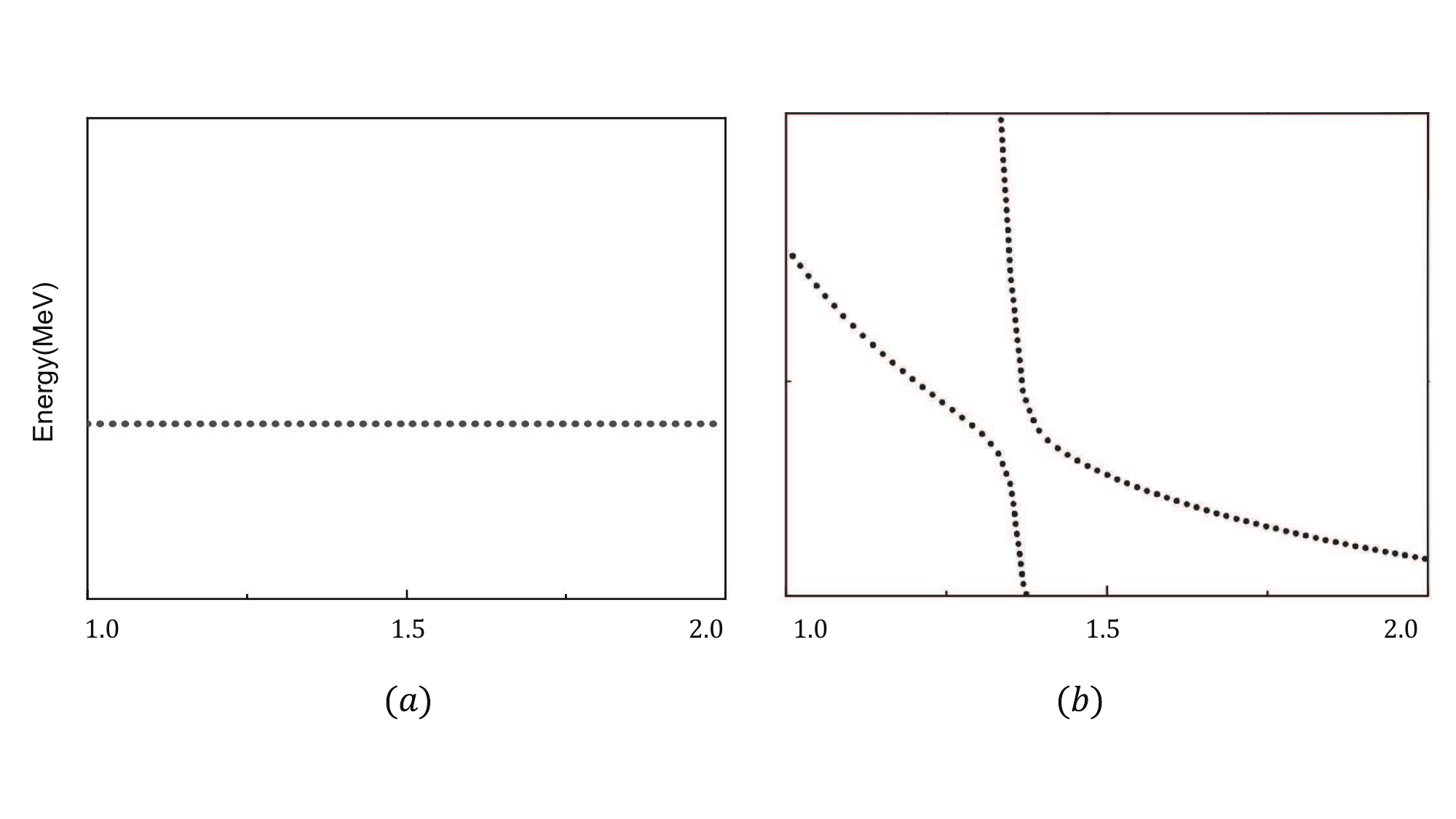}}
  \caption{Two forms of resonant states (a) the resonance has weak coupling (or no coupling) with the scattering states; (b) the resonances has strong coupling with the scattering states;}
\label{RSM}
\end{figure}

\begin{equation}\label{formula_RSM}
\Gamma =4 |V(\alpha_c)|\frac{\sqrt{|S_r||S_s|}}{|S_r-S_s|}
\end{equation}

\section{Results and discussions}

\begin{table}[!t]
\caption{\label{Boundstate} Results of the bound state calculations in the $c\bar{c}c\bar{c}$ system.(unit: MeV)}
\begin{ruledtabular}
\begin{tabular}{ccccc}
Channel             &$|[LS]_i F_j C_k\rangle$  & E       & Mixed~~ \\
$\eta_c \eta_c$     &$|111\rangle$             & $5962$  &5961\\
$J/\psi J/\psi$     &$|321\rangle$             & $6201$  \\
$\chi_{c0}\chi_{c0}$&$|221\rangle$             & $6858$  \\
$\chi_{c1}\chi_{c1}$&$|521\rangle$             & $6988$  \\
$\chi_{c2}\chi_{c2}$&$|712\rangle$             & $7089$  \\
$\chi_{c1}h_c$      &$|612\rangle$             & $7004$  \\
$h_c h_c$           &$|412\rangle$             & $7018$  \\
\\
 $[\eta_c]_8[\eta_c]_8$& $|112\rangle$ &6454 & 6300\\
 $[J/\psi]_8[J/\psi]_8$& $|222\rangle$ &6430 & \\
 $[cc]_6^0[\bar{c}\bar{c}]_{\bar{6}}^0$ & $|134\rangle$ &6445 & \\
 $[cc]_3^1[\bar{c}\bar{c}]_{\bar{3}}^1$ & $|243\rangle$ &6411 & \\

\multicolumn{3}{c}{Complete coupled-channels:}  & $5960$   \\
\end{tabular}
\end{ruledtabular}
\end{table}

In this section, we focus on the identification of genuine resonance states. Within the quark model, two mechanisms contribute to the formation of resonance states. The first mechanism involves  meson exchange within a color-singlet molecular state. The second mechanism arises from the attractive forces between different colorful structures, including color octet and diquark strcuture.  Subsequently, considering the the attractive forces between different structures, we performed channel-coupling for all colorful channels. In this calculation, the obtained energy levels are theoretically resonances. Among these, some are false states generated due to limitations in theoretical calculation space. Therefore, the stability of these energy levels is subsequently examined by the Real-Scaling method. Due to the experimentally observed resonances in $c\bar{c}c\bar{c}$ system, such as $X(6900)$ and $X(7200)$, having energy levels much higher than the lowest threshold channel $\eta_c \eta_c$, in our calculations, we additionally considered more excited states, such as $\chi_{c0}\chi_{c0}$, $\chi_{c1}\chi_{c1}$, $\chi_{c2}\chi_{c2}$, $\chi_{c1}h_c$, and $h_ch_c$.

\begin{table}[!t]
\caption{\label{Resonantstate}Results of the resonance calculations in the $c\bar{c}c\bar{c}$ system. (unit: MeV)}
\begin{ruledtabular}
\begin{tabular}{ccccc}
  ~~Resonances~~& $[cc]_3^1[\bar{c}\bar{c}]_{\bar{3}}^1$ &$[cc]_6^0[\bar{c}\bar{c}]_{\bar{6}}^0$ & $[\eta_c]_8[\eta_c]_8$   & $[J/\psi]_8[J/\psi]_8$\\
 ~~$R(6300)$~~  &   55.0\% ~~ &    0.1\%  ~~&~35.5\% ~~ & ~~9.4\%  ~~\\
 ~~$R(6390)$~~  &   54.6\%~~  & ~~ 0.1\%  ~~& 37.8\% ~~ &   7.6\%~~\\
 ~~$R(6610)$~~  & ~~ 1.8\%~~  & ~~41.9\%  ~~& 10.1\% ~~ &  46.1\%~~\\
 ~~$R(6690)$~~  & ~~ 0.1\%~~  & ~~42.3\%  ~~& 17.2\% ~~ &  39.8\%~~\\
 ~~$R(6850)$~~  & ~~ 0.1\%~~  & ~~42.2\%  ~~& 17.9\% ~~ &  40.5\%~~\\
 ~~$R(6920)$~~  & ~~ 0.2\%~~  & ~~42.3\%  ~~& 15.4\% ~~ &  42.1\%~~\\
 ~~$R(7000)$~~  & ~~20.5\%~~  & ~~23.6\%  ~~& 47.4\% ~~ &  8.5\%~~\\
 ~~$R(7080)$~~  & ~~43.6\%~~  & ~~ 6.6\%  ~~& 49.7\% ~~ &  0.1\%~~\\
 ~~$R(7160)$~~  & ~~55.4\%~~  & ~~ 0.6\%  ~~& 41.4\% ~~ &  4.5\%~~\\
 ~~$R(7210)$~~  & ~~44.2\%~~  & ~~ 6.3\%  ~~& 49.5\% ~~ &  0.1\%~~\\
 ~~$R(7280)$~~  & ~~55.4\%~~  & ~~ 0.6\%  ~~& 30.6\% ~~ & 13.5\%~~\\
\end{tabular}
\end{ruledtabular}
\end{table}

\textbf{Bound-state calculation:} TABLE \ref{Boundstate} exhibits the results of the bound-state calculations. Concerning the color-singlet structure, there are two ground states, specifically $\eta_c \eta_c$ and $J/\psi J/\psi$, accompanied by five highly-excited color singlets: $\chi_{c0} \chi_{c0}$, $\chi_{c1} \chi_{c1}$, $\chi_{c2} \chi_{c2}$, $\chi_{c1} h_{c}$, and $h_{c} h_{c}$. The energy of the ground state is relatively low, around 6.0 GeV, while the energy of the highly excited state is in the range of 6.85-7.02 GeV. The coupling of these seven color-singlet channels did not result in a bound state but only pushed the lowest energy of system down by 1 MeV, still higher than the energy of the lowest threshold $\eta_c \eta_c$. For the color-excited state structure, we only consider the color octet of two ground states, $[\eta_c]_8 [\eta_c]_8$ and $[J/\psi]_8[ J/\psi]_8$, and the diquark of two ground states, $[cc]_6^0[\bar{c}\bar{c}]_{\bar{6}}^0$, $[cc]_3^1[\bar{c}\bar{c}]_{\bar{3}}^1$. Given that the quarks in the $c\bar{c}c\bar{c}$ system are heavy quarks, the quark separations in both the color octet and diquark structures are quite small, roughly within 0.5 fm. Consequently, their energies are close, around the range of 6.4 GeV. As a result, the color structure coupling effect is so strong that the minimum energy is reduced to 6.3 GeV. Nevertheless, the complete-coupling of all channels of the colorful structure and all channels of the color-singlet structure causes the lowest energy $\eta_{c} \eta_{c}$ energy to be depressed by 1 MeV, still above the threshold energy.  This implies that, in our calculations, there are no bound states in the $c\bar{c}c\bar{c}$ system. It also indicates that even if resonant states exist, their primary components are unlikely to be molecular states.

\begin{figure}[!th]
\vspace{-0.2cm}
\setlength {\abovecaptionskip} {-0.6cm}
\setlength {\belowcaptionskip} {-0.2cm}
  \centering
  \includegraphics[width=9cm,height=8cm]{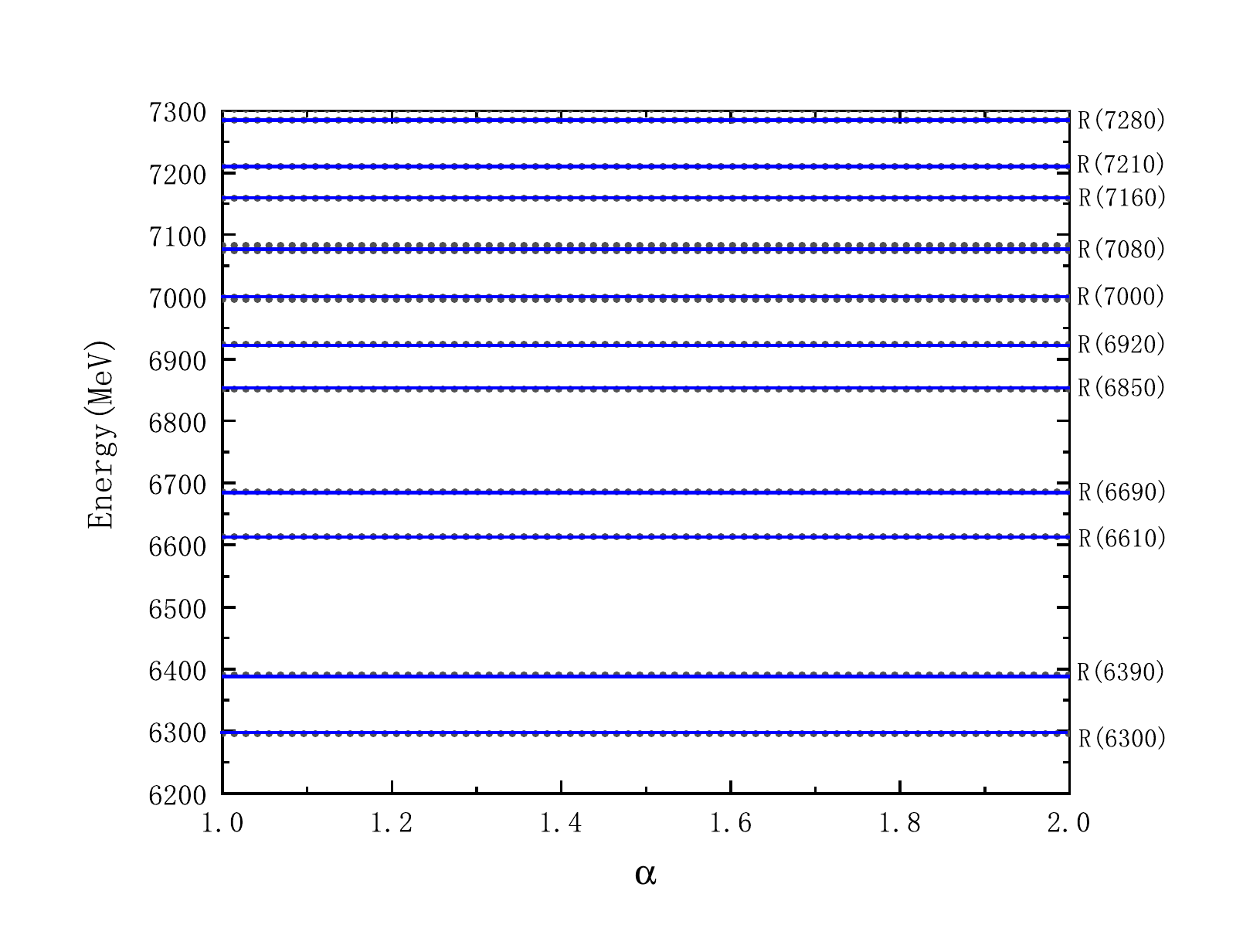}
  \caption{Energy spectrum by only calculating four colorful channels.}
   \label{Resonances}
\end{figure}
\begin{figure}[!h]
\vspace{-0.2cm}
\setlength {\abovecaptionskip} {-0.6cm}
\setlength {\belowcaptionskip} {-0.2cm}
  \centering
  \includegraphics[width=9cm,height=12cm]{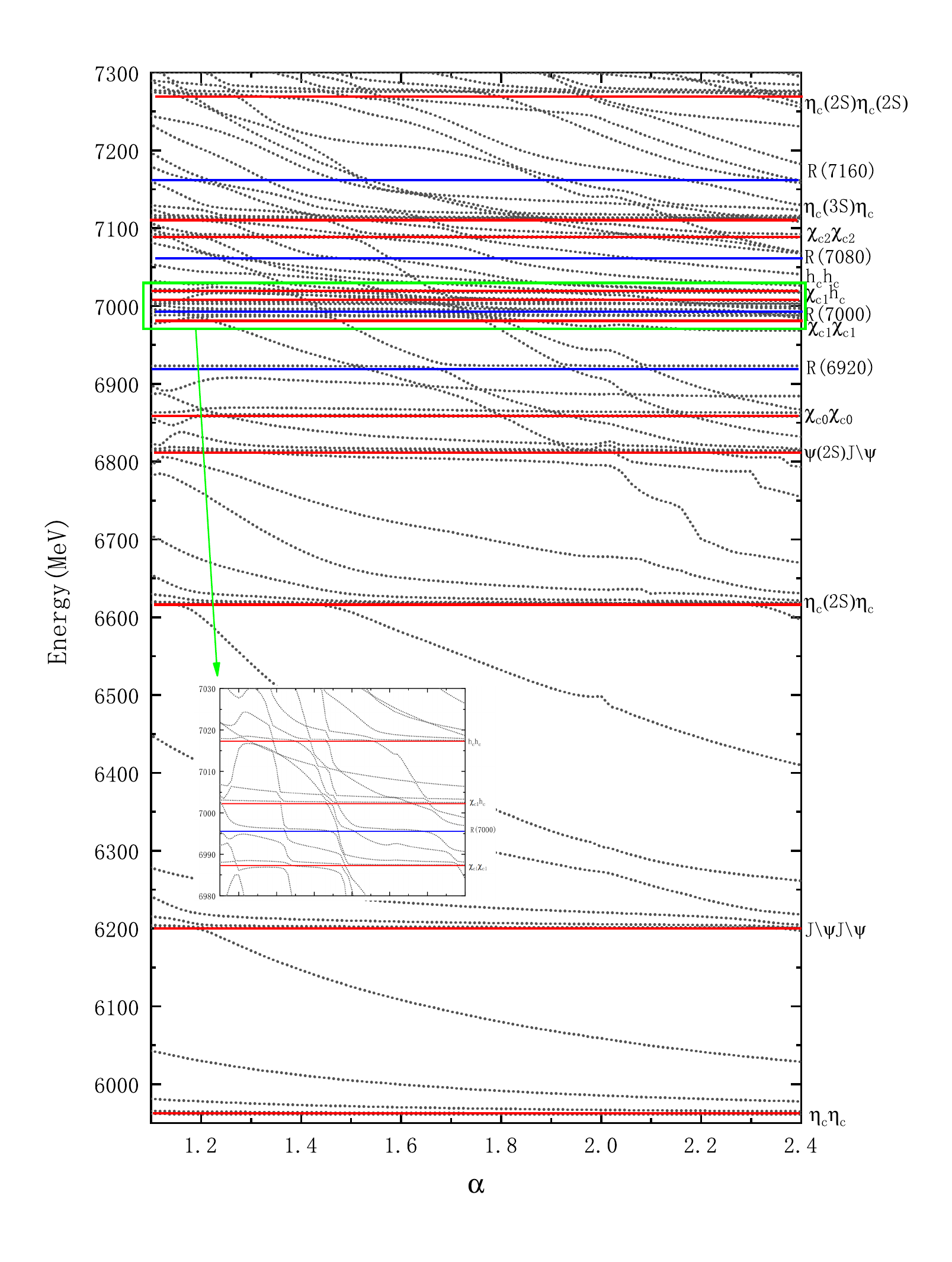}
  \caption{Energy spectrum by complete-coupling channels. The blue line depicts the resonant state, while the red line represents the physical threshold.}
   \label{Resonances_12}
\end{figure}

\textbf{Resonance-state calculation:} Since the bound-state calculation results indicate an absence of bound states in the $c\bar{c}c\bar{c}$ system, we next consider to search for resonances in the colorful structure (color octet and diquark structure) and use the real-scaling method to find stable states.

\begin{figure*}[htbp]
\centering

\subfigure[Resonances only decay to $\eta_c \eta_c$]{
\begin{minipage}[t]{0.5\linewidth}
\centering
\includegraphics[width=6cm,height=6cm]{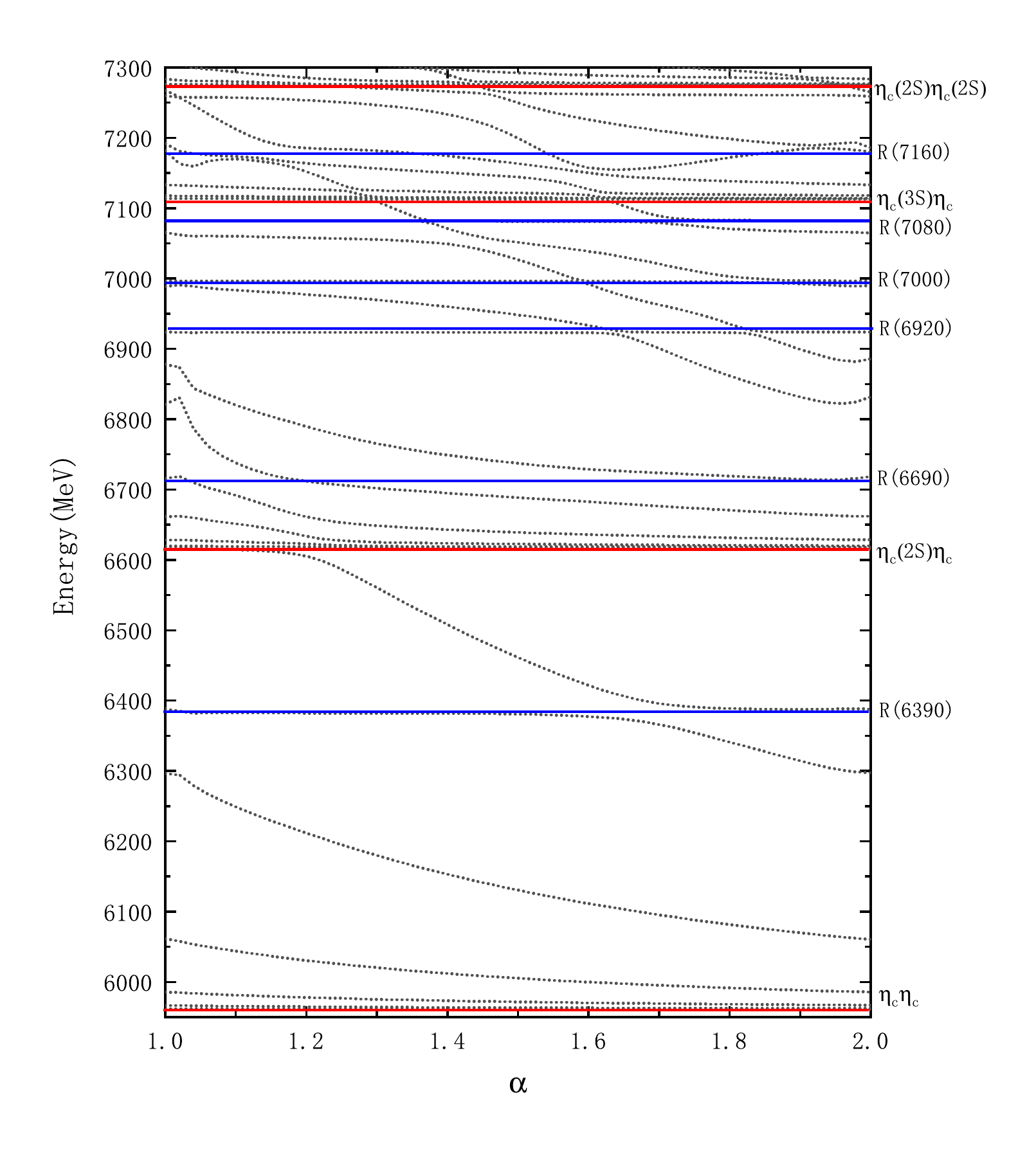}
\end{minipage}%
}%
\subfigure[Resonances only decay to $J/\psi J/\psi$]{
\begin{minipage}[t]{0.5\linewidth}
\centering
\includegraphics[width=6cm,height=6cm]{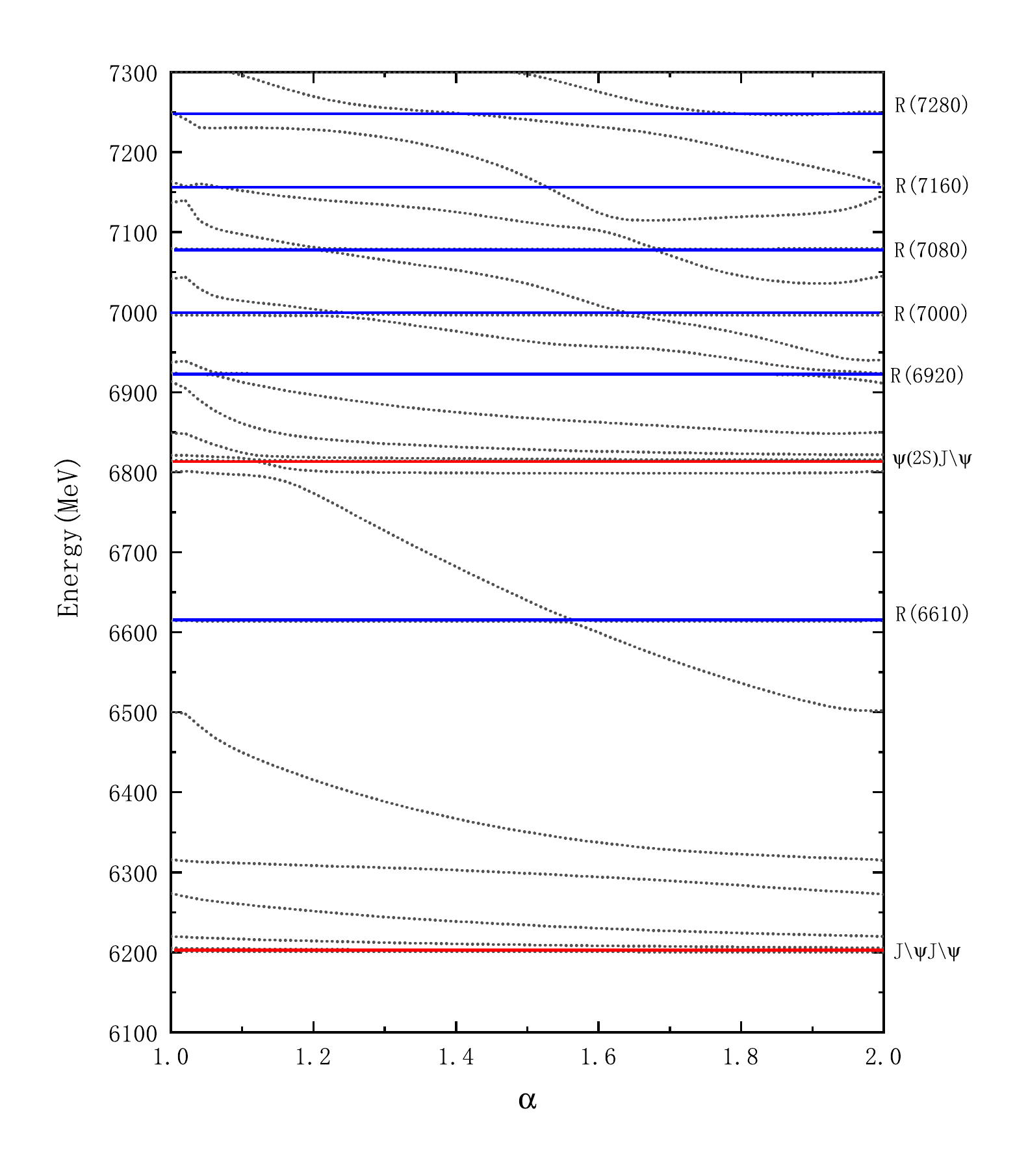}
\end{minipage}%
}

\subfigure[Resonances only decay to $\chi_{c0}\chi_{c0}$]{
\begin{minipage}[t]{0.3\linewidth}
\centering
\includegraphics[width=6cm,height=6cm]{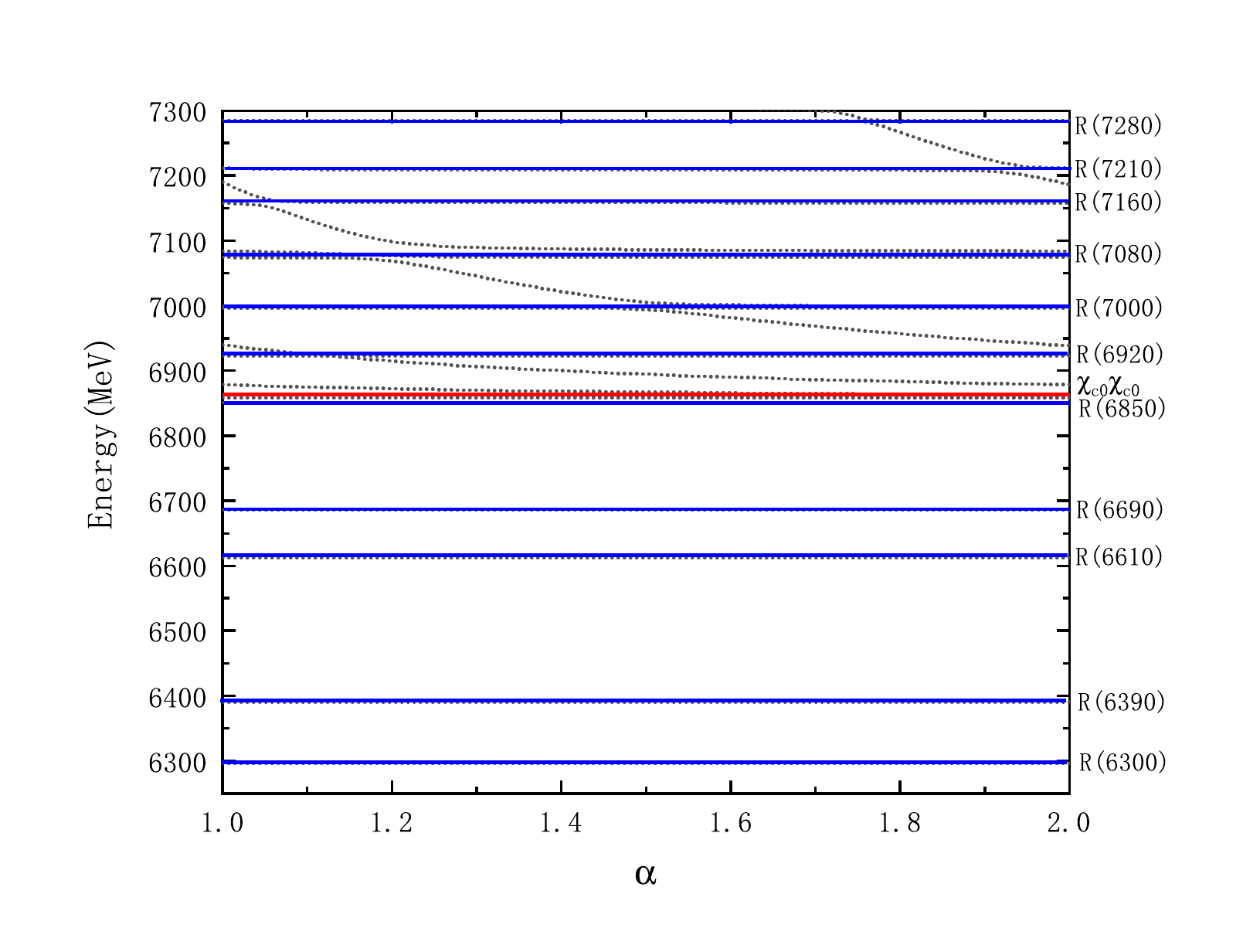}
\end{minipage}
}
\subfigure[Resonances only decay to $\chi_{c1}\chi_{c1}$]{
\begin{minipage}[t]{0.3\linewidth}
\centering
\includegraphics[width=6cm,height=6cm]{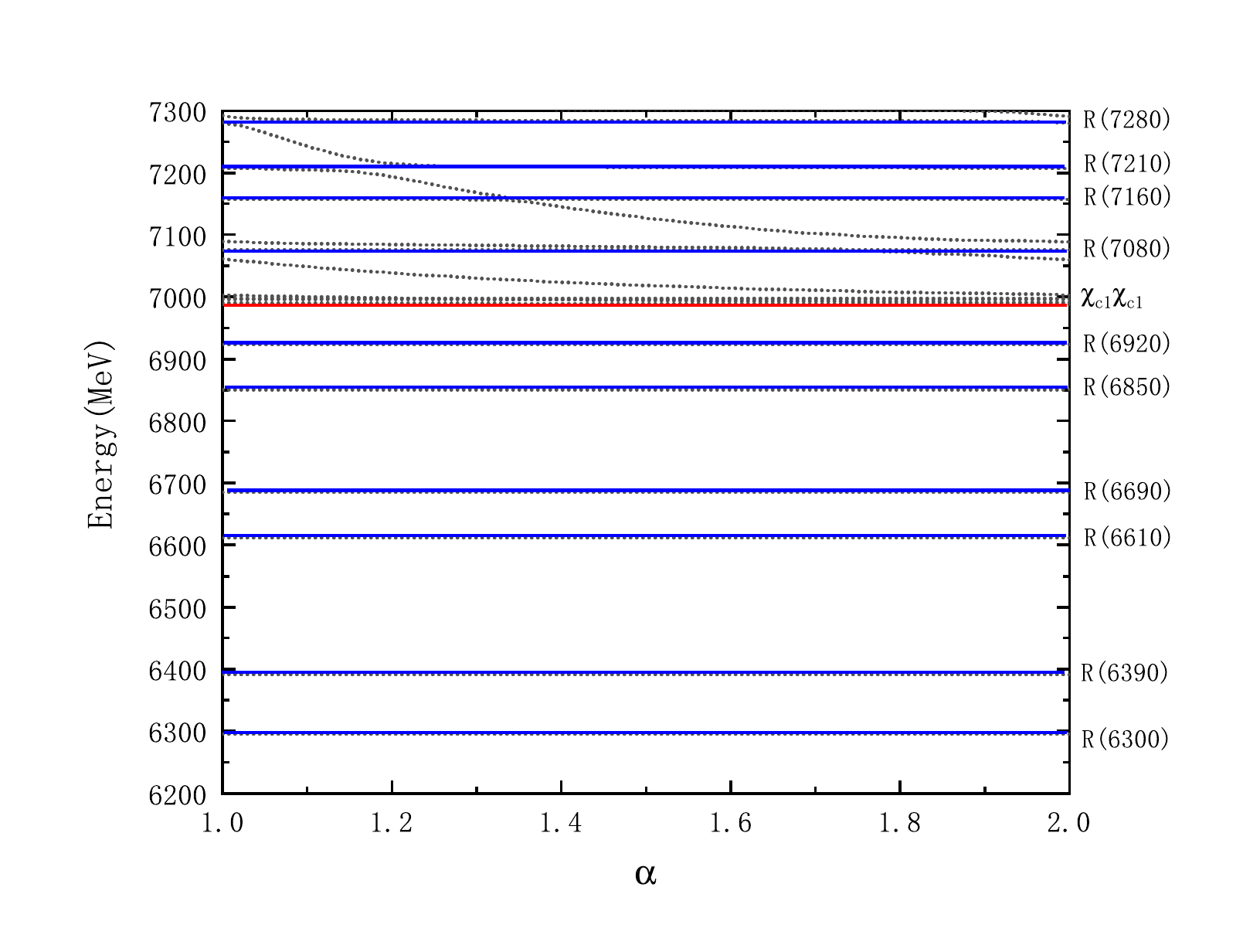}
\end{minipage}
}
\subfigure[Resonances only decay to $\chi_{c2}\chi_{c2}$]{
\begin{minipage}[t]{0.3\linewidth}
\centering
\includegraphics[width=6cm,height=6cm]{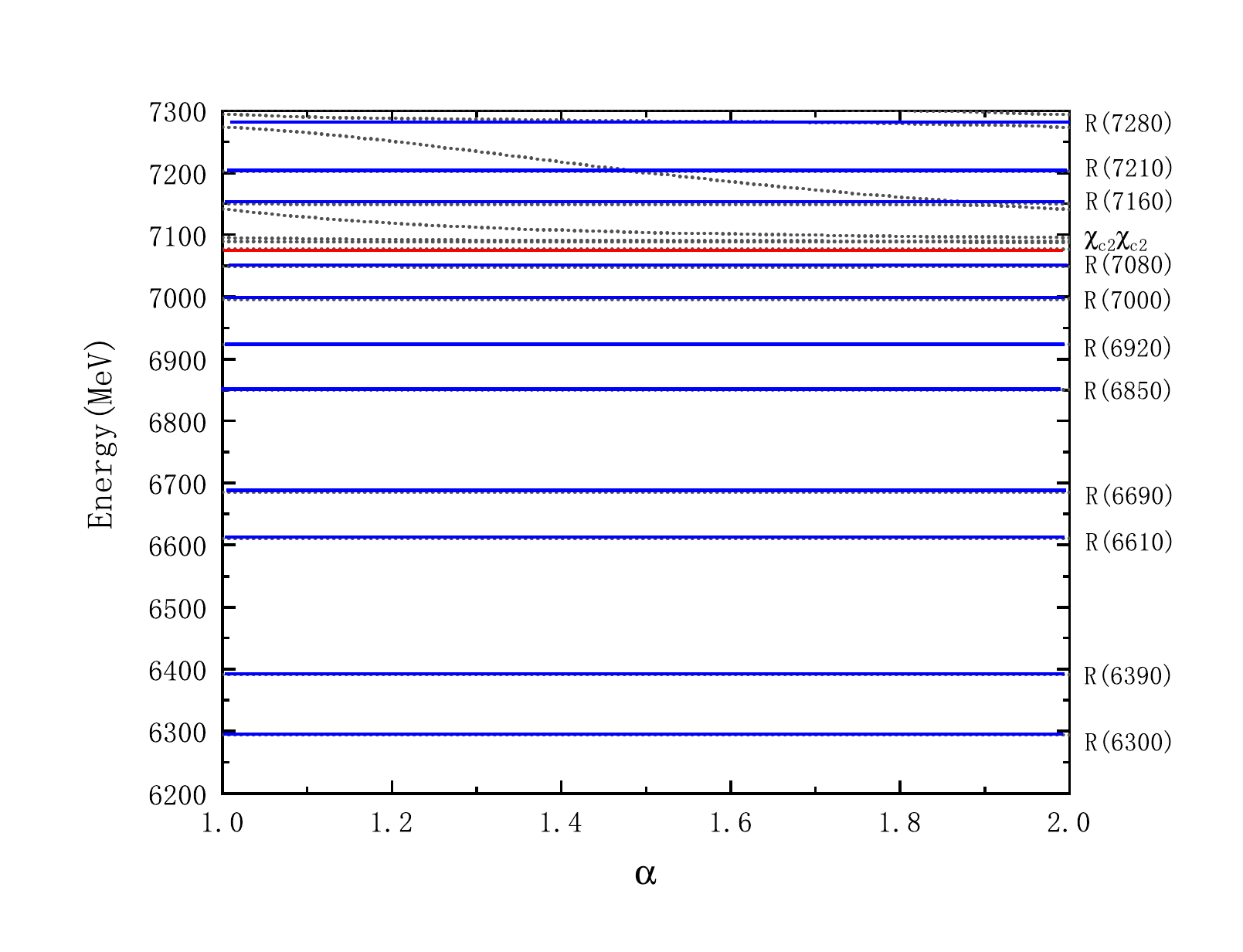}
\end{minipage}
}
\subfigure[Resonances only decay to $\chi_{c1}h_c$]{
\begin{minipage}[t]{0.3\linewidth}
\centering
\includegraphics[width=6cm,height=6cm]{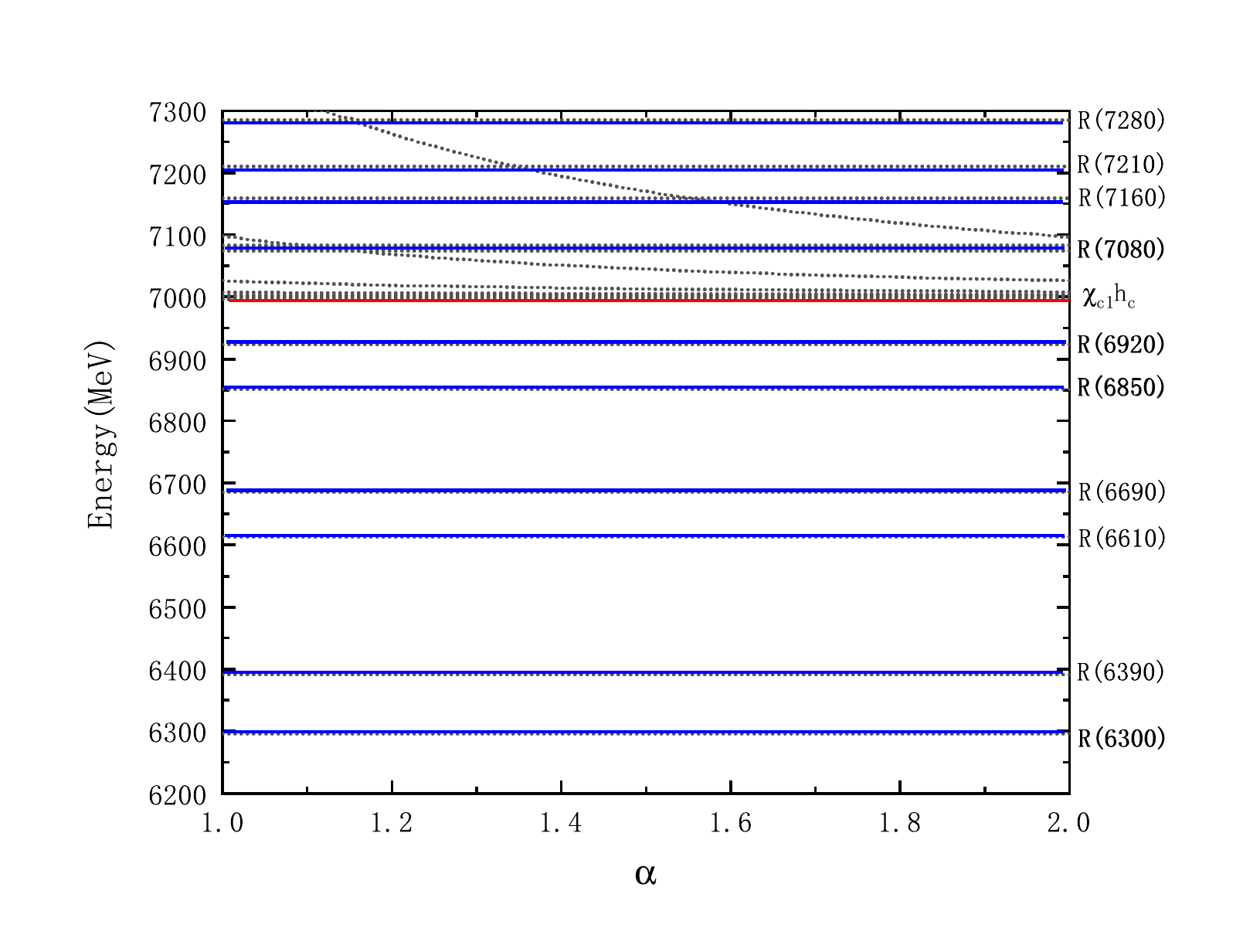}
\end{minipage}
}%
\subfigure[Resonances only decay to $h_{c}h_c$]{
\begin{minipage}[t]{0.3\linewidth}
\centering
\includegraphics[width=6cm,height=6cm]{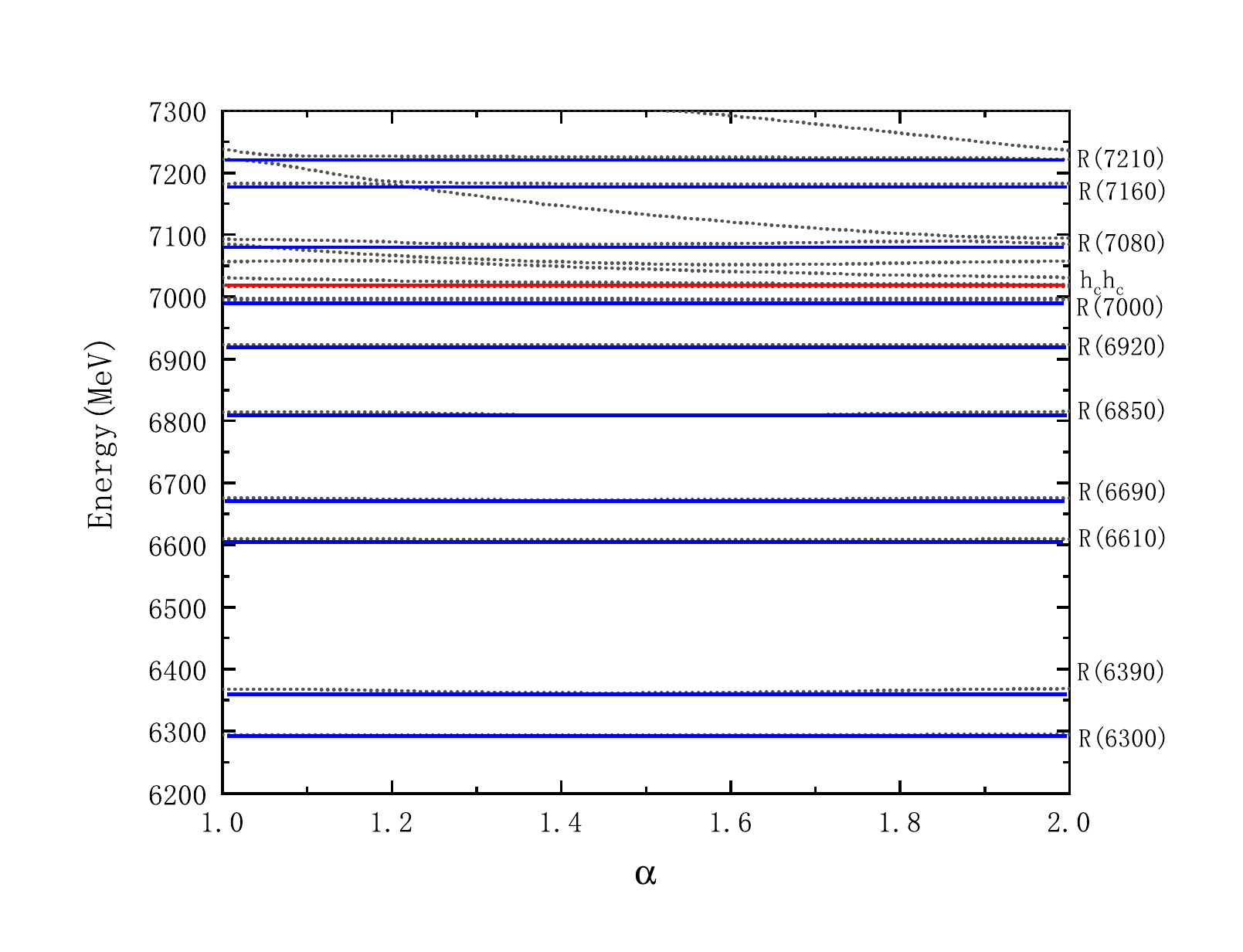}
\end{minipage}
}%

\centering
\caption{ Real-scaling diagrams illustrating the decay of resonances to different threshold channels in the $0^{+}$ $c\bar{c}c\bar{c}$ system.}\label{Fic}
\end{figure*}
Firstly, we coupled the four colorful structures, including $[cc]_3^1[\bar{c}\bar{c}]_{\bar{3}}^1$, $[cc]_6^0[\bar{c}\bar{c}]_{\bar{6}}^0$, $[\eta_c]_8[\eta_c]_8$, and $[J/\psi]_8[J/\psi]_8$, to obtain the theoretical resonances. Considering the experimentally observed resonances ($X(6400)$, $X(6600)$, $X(6900)$, and $X(7200)$) with energy ranges between 6.4 GeV and 7.3 GeV, we selected a total of 11 energy levels within the range of 6.2 GeV to 7.3 GeV in the color channel coupling results, denoted as $R(energy)$, shown in FIG. \ref{Resonances}. We have analyzed the main components of these states, and the results are listed in TABLE \ref{Resonantstate}. According to the results of the current experiment, there are four possible resonance states, which are $X(6400)$, $X(6600)$, $X(6900)$, and $X(7200)$. In our results, we identified candidates corresponding to the observed states in the experiment: $R(6390)$ serves as the candidate for $X(6400)$, $R(6610)$ is considered a candidate for $X(6600)$, $R(6920)$ is  a candidate for $X(6900)$, and $R(7160)$, $R(7210)$, and $R(7280)$ collectively stand as candidates for $X(7200)$. We then validated the stability of these candidates by real-scaling method. The energy of these obtained resonance states remains unchanged with varying $\alpha$ factors, indicating their stability. Subsequently, all possible decay channels are considered to check whether these possible resonances can survive coupling with decay channels. As illustrated in FIG. \ref{Resonances_12}, the most of the calculated resonances decay, leaving only four genuine resonances: $R(6920)$, $R(7000)$, $R(7080)$, and $R(7160)$.

In order to understand the decay processes of the other unstable resonances and simultaneously calculate the decay widths of these genuine resonances, we calculated the real-scaling diagrams for the coupling of each decay channel with the resonances, as shown in FIG. \ref{Fic}. FIG. \ref{Fic} (a) illustrates the coupling results of all resonances with $\eta_c \eta_c$. Because $\eta_c(2S) \eta_c$ and $\eta_c(2S) \eta_c(2S)$ are so close in energy to resonances $R(6610)$ and $R(7280)$ in our calculation, it is clear that a very strong coupling effect results, which causes these two resonances ( $R(6610)$ and $R(7280)$ ) to decay very quickly. For the same reasons, $R(6300)$, $R(6850)$, and $R(7210)$ did not survive in our calculations. In summary, among the 11 theoretical resonances in FIG. \ref{Resonances}, $R(6300)$, $R(6610)$, $R(6850)$, $R(7210)$, and $R(7280)$ will decay into the $\eta_c \eta_c$ decay channel.
Further considering all resonances with the ground decay channel $J/\psi J/\psi$, as shown in FIG. \ref{Fic} (b), a total of 6 resonant states survived: $R(6610)$, $R(6920)$, $R(7000)$, $R(7080)$, $R(7160)$, and $R(7280)$. It can be observed that in FIG. \ref{Fic} (a), $R(6390)$ and $R(6690)$ decay into the $J/\psi J/\psi$ channel, which indicates our calculations do not support the experimentally observed $X(6400)$ and $X(6600)$.  Considering both the $\eta_c \eta_c$ and $J/\psi J/\psi$ ground-state decay channels, the genuine resonances are only $R(6920)$, $R(7000)$, $R(7080)$ and $R(7160)$.  FIG \ref{Fic} (c, d, e, f, g) show the decays of all resonances into five excited states: $\chi_{c0} \chi_{c0}$, $\chi_{c1} \chi_{c1}$, $\chi_{c2} \chi_{c2}$, $\chi_{c1} h_{c}$, and $h_{c} h_{c}$. Due to their significantly higher energies compared to the ground states, the four stable resonances $R(6920)$, $R(7000)$, $R(7080)$, and $R(7160)$ remain stable.

\begin{table}[tp]
\centering
\caption{Various decay channels and corresponding decay widths of the obtained resonances. (unit: MeV)\label{Width}}
\begin{tabular}{cccccccccc}
\hline \hline
Decay channels~~~~ & $R(6920)$ &$R(7160)$&  $R(7000)$ &$R(7080)$\\
\hline
$\eta_c\eta_c$      &  1.1      &  9.3  &  7.9      &  12.4     \\
$J/\psi J/\psi$     &  9.8      &  22.9 &  10.3     &  1.9     \\
$\chi_{c0}\chi_{c0}$&  0.2      &  35.1 &  26.2     & 34.8     \\
$\chi_{c1}\chi_{c1}$&   -       &  2.7  &  0.2      &  1.8    \\
$\chi_{c2}\chi_{c2}$&   -       &  2.5  &   -       &   -    \\
$\chi_{c1}h_c$      &   -       &  0.4  &   -       &  0.3    \\
$h_{c}h_c$          &   -       &  4.9   &   -       &  8.8    \\
Total               &  10.1     &  77.8 &  44.6     &  60.0    \\
\hline
\end{tabular}
\end{table}

In TABLE \ref{Width}, four resonance states are listed, and we analyze the potential decay channels for each resonance state. Concerning $R(6920)$, identified as the candidate for $X(6900)$ in experiments, three possible decay channels exist: $\eta_c\eta_c$, $J/\psi J/\psi$, and $\chi_{c0} \chi_{c0}$. Among these, $J/\psi J/\psi$ serves as the dominant decay channel, contributing to over 90\% of the total decay width. It is worth noting that the total width of $R(6920)$ is much smaller than the experimental value. The candidate $R(7160)$ for the experimentally observed  $X(7200)$ exhibits two primary decay channels: $J/\psi J/\psi$ and $\chi_{c0} \chi_{c0}$. The corresponding decay widths are 22.9 MeV and 35.1 MeV, respectively. The decay widths for other decay channels are comparatively smaller. Considering that the experimental $X(7200)$ is predominantly observed in the $J/\psi J/\psi$ channel, we propose an experimental search for $X(7200)$ in the $\chi_{c0} \chi_{c0}$ channel. We also predict two other resonance states, $R(7000)$ and $R(7080)$, which have similar decay channels, mainly in the $\chi_{c0} \chi_{c0}$. Because the energy of $R(7080)$ is higher, its decay phase space is larger, so its decay width is larger than that of $R(7000)$, about 60 MeV.

\section{Summary}

In the framework of chiral quark model, we systematically study the $c\bar{c}c\bar{c}$ system of $J^{P}=0^{+}$ by using multi-Gaussian expansion method. We consider not only the diquark structure of space orbit as ground state, but also the dimeson structure of various space as excited state.

The bound state calculation shows that although the contributions of various excited states are taken into account, their coupling effects do not cause bound states to appear in the $c\bar{c}c\bar{c}$ system. Then, we used the Real-scaling method to find the resonant states in the $c\bar{c}c\bar{c}$ system. Due to color attraction, the color structure includes color octet and diquark structure in which every energy level is a resonance state. The color structure is structurally coupled to all color singlets, and the stability of the resonant states is tested using the $\alpha$factor scaling space. In this case, we obtain four resonant states, $R(6920)$,$R(7000)$,$R(7080)$ and $R(7160)$.

In our calculation, $R(6920)$ is a strong candidate for $X(6900)$ on experiment, and $R(7160)$ is a strong candidate for $X(7200)$ on experiment. Their decay widths show that $R(7160)$ has two main decay channels, and in addition to $J/\psi J/\psi$, $\chi_{c0} \chi_{c0}$ is also the main decay channel. The other two we predicted resonance, $R(7000)$, $R(7080)$ the main decay way is $\chi_{c0} \chi_{c0}$. Therefore, we recommend experiments to look for them in the future.

\acknowledgments{This work is supported partly by the National Natural Science Foundation of China under Grant Nos. 12205249, 11675080, 11775118 and 11535005,  and the Funding for School-Level Research Projects of Yancheng Institute of Technology (No. xjr2022039).}

\end{document}